\documentclass[reprint, aps, prd, nofootinbib, superscriptaddress, showkeys]{revtex4-2}
\usepackage{graphicx}
\usepackage{amsmath}
\usepackage{amssymb}
\usepackage{bm}

\begin{document}

\title{Search for the radiative decay of the cosmic neutrino background through spectral measurements of the cosmic infrared background using PRIMA}

\author{Yuji Takeuchi}
\email{takeuchi@hep.px.tsukuba.ac.jp}
\affiliation{University of Tsukuba, Division of Physics and Tomonaga Center for the History of the Universe, Faculty of Pure and Applied Sciences, 1-1-1 Tennodai, Tsukuba, Japan, 305-8571}

\author{Shuji Matsuura}
\affiliation{Kwansei Gakuin University, School of Science, Department of Physics and Astronomy, 2-1 Gakuen, Sanda, Japan, 669-1337}

\author{Shinhong Kim}
\affiliation{University of Tsukuba, Division of Physics and Tomonaga Center for the History of the Universe, Faculty of Pure and Applied Sciences, 1-1-1 Tennodai, Tsukuba, Japan, 305-8571}

\author{Takashi Iida}
\affiliation{University of Tsukuba, Division of Physics and Tomonaga Center for the History of the Universe, Faculty of Pure and Applied Sciences, 1-1-1 Tennodai, Tsukuba, Japan, 305-8571}

\date{\today}

\begin{abstract}
We propose to search for a faint yet distinguishable contribution to the cosmic infrared background (CIB) spectrum arising from the radiative decay of the cosmic neutrino background (C$\nu$B). In the Standard Model of particle physics, neutrino decay is highly suppressed, with a predicted lifetime on the order of $10^{43}$~years. However, non-standard models suggest the possibility of significantly shorter lifetimes, ranging from $10^{12}$ to $10^{17}$ years. Observations to date, however, only provide a lower limit of approximately $10^{12}$~years for the neutrino lifetime. In PRIMA's low-resolution mode ($R \sim 100$), a diffuse background analysis, combined with the removal of point sources associated with known galaxies in a wide-field ($\sim 1$~square degree) spectroscopic survey covering in the 24--240~$\mu$m range could facilitate a search for neutrino decay lifetimes up to O($10^{15}$) years. The expected signal from C$\nu$B decay at 50~$\mu$m for a neutrino lifetime of O($10^{15}$) years is more than an order of magnitude fainter than the CIB and over three orders of magnitude fainter than the zodiacal emission foreground. Taking advantage of the characteristic spectral features of C$\nu$B decay, this level of sensitivity can be achieved with approximately 100 hours of total observation time, based on estimated surface brightness sensitivity. Searching for neutrino decay at this sensitivity could place strong constraints on several non-standard theories. A positive detection would provide compelling evidence for non-standard contributions to neutrino decay and could directly reveal the cosmic neutrino background. Furthermore, the decay photon spectrum could offer insights into the absolute mass of neutrinos, and key cosmological parameters.
\end{abstract}


\maketitle

\section{Introduction}

A few seconds after the birth of the universe, neutrinos became thermally decoupled from other particles. The neutrinos present at that time have since persisted almost uniformly throughout the universe, forming what is known as the cosmic neutrino background (c$\nu$B). Based on measurements of the cosmic microwave background (CMB), the c$\nu$B is predicted to have a temperature of 1.95~K and a density of approximately 110 particles per $\mathrm{cm}^3$ per generation. While indirect evidence for c$\nu$B has been obtained through CMB measurements\cite{ref1}, direct detection remains unachieved.

Neutrinos are known to have three mass eigenstates, with the flavor eigenstates being a mixture of these mass eigenstates, as established by oscillation experiments\cite{ref2,ref3}. The mass-squared differences between the mass eigenstates have been measured, making it theoretically possible for the heavier neutrino to decay into a lighter neutrino accompanied by the emission of a single photon ($\nu_3 \rightarrow \nu_{1,2}+\gamma$). However, within the Standard Model of particle physics, neutrino decay processes are subject to a very strong suppression mechanism, resulting in a predicted lifetime for the heaviest neutrino ($\nu_3$) on the order of $10^{43}$~years\cite{ref4}. Conversely, the neutrino decay process provides an exceptionally sensitive test of the Standard Model. Certain non-standard models, such as those proposed in Refs.\cite{ref5} and \cite{ref6}, predict that this lifetime of $\nu_3$ could be significantly shorter, ranging from $10^{12}$ to $10^{17}$~years, depending on the allowed range of parameters in these models. Observationally, the lifetime of $\nu_{3}$ is constrained only by a lower bound of approximately $10^{12}$~years\cite{ref7}, derived from measurements of the cosmic infrared background radiation. In this study, we aim to improve upon this lower limit by searching for potential radiative decay signatures of the cosmic neutrino background in the cosmic infrared background. Our analysis is expected to achieve a sensitivity more than two orders of magnitude beyond the current limit, providing tighter constraints on both the Standard Model and possible non-standard neutrino decay scenarios.

\section{Expected Spectrum from the Radiative Decay of the Cosmic Neutrino Background}
\label{sec2}

Since neutrino radiative decay is a two-body decay, the photon energy observed in the rest frame of the parent particle is monoenergetic, as expressed in Eq.~\eqref{eqn1}.

\begin{equation}
\label{eqn1}
\lambda_{0} = \frac{h}{c}\frac{2m_{3}}{m_{3}^{2} - m_{1,2}^{2}}~.
\end{equation}

In the case of neutrino radiative decay ($\nu_{3} \rightarrow \nu_{1,2}+\gamma$), $\lambda_{0}$ is the emitted photon wavelength, $m_{3}$ is the mass of the decaying neutrino, and $m_{1,2}$ represents the mass of the lighter neutrino state.

The wavelength and luminosity of the decay photons produced by neutrino decay at redshift $z$, when observed from Earth's vicinity, are affected by the $z$-dependence of the density and decay rate of the c$\nu$B, as well as the redshift of the decay photons itself. Additionally, they are influenced by the Doppler shift due to the thermal motion of the c$\nu$B, whose temperature also depends on $z$. Assuming no influence from the thermal motion of neutrinos, the wavelength dependence of the surface brightness of the c$\nu$B decay photons observed from Earth's vicinity can be expressed as:

\begin{equation}
\label{eqn2}
\lambda I_{\lambda} = \frac{\rho_{\nu,0}}{4\pi\tau_{3}}\frac{hc^{2}}{\lambda H_{0}}\left\lbrack \left( \frac{\lambda}{\lambda_{0}} \right)^{3}\Omega_{M,0} + \Omega_{\Lambda,0} \right\rbrack^{- \frac{1}{2}}~.
\end{equation}

Here, $\lambda$ is the observed wavelength of the decay photons, $\tau_{3}$ is the lifetime of $\nu_{3}$, $\rho_{\nu,0}$ is the current neutrino density, $H_{0}$ is the Hubble constant, $\lambda_{0}$ is the decay photon wavelength in the $\nu_{3}$ rest frame, and $\Omega_{M,0}$ and $\Omega_{\Lambda,0}$ represent the current matter density and cosmological constant, respectively. In this article, the predictions for surface brightness are based on $\rho_{\nu,0} = 110/\rm{cm}^{3}$, $H_{0} = 70~\rm{km/s/Mpc}$, $\Omega_{M,0} = 0.3$, and $\Omega_{\Lambda,0} = 0.7$. The following equation describes the redshift-dependent c$\nu$B temperature and the corresponding relativistic Fermi-Dirac velocity distribution for neutrinos:

\begin{equation}
\label{eqn3}
f\left( p,z \right) = \frac{1}{\exp\left\lbrack \frac{pc}{kT_{0}\left( 1 + z \right)} \right\rbrack + 1}~.
\end{equation}

Here $p$ represents the momentum of neutrinos, and $T_{0}$ denotes the current c$\nu$B temperature, taken as 1.95~K. To account for the influence of neutrino thermal motion, we use the distribution shown in Eq.~\eqref{eqn3} to generate isotropic thermal motion of neutrinos via a Monte Carlo method for each $\lambda$-bin in Eq.~\eqref{eqn2}. The resulting distribution from Eq.~\eqref{eqn2} is then convolved with the Doppler shift effects caused by the line-of-sight velocities.

Assuming a mass of 50~meV for $\nu_{3}$, the energy of the emitted photon would be approximately 25~meV (corresponding to a wavelength of 50~$\mu$m). An upper limit of 230~meV for the sum of the neutrino masses has been established through cosmic observations\cite{ref8}. Using the mass-squared differences measured in neutrino oscillation experiments, the mass of $\nu_{3}$ is expected to be in the range of 50--87~meV, and the wavelength of the decay photons in the rest frame of $\nu_{3}$ is predicted to fall between 51 and 89~$\mu$m. Given these conditions, exploring the decay photons from the c$\nu$B in the 50~$\mu$m wavelength range, with a sensitivity exceeding the current lower limit of $10^{12}$~years, holds significant importance for both cosmology and particle physics.

\begin{figure}[htbp]
  \centering
  \includegraphics[width=\columnwidth]{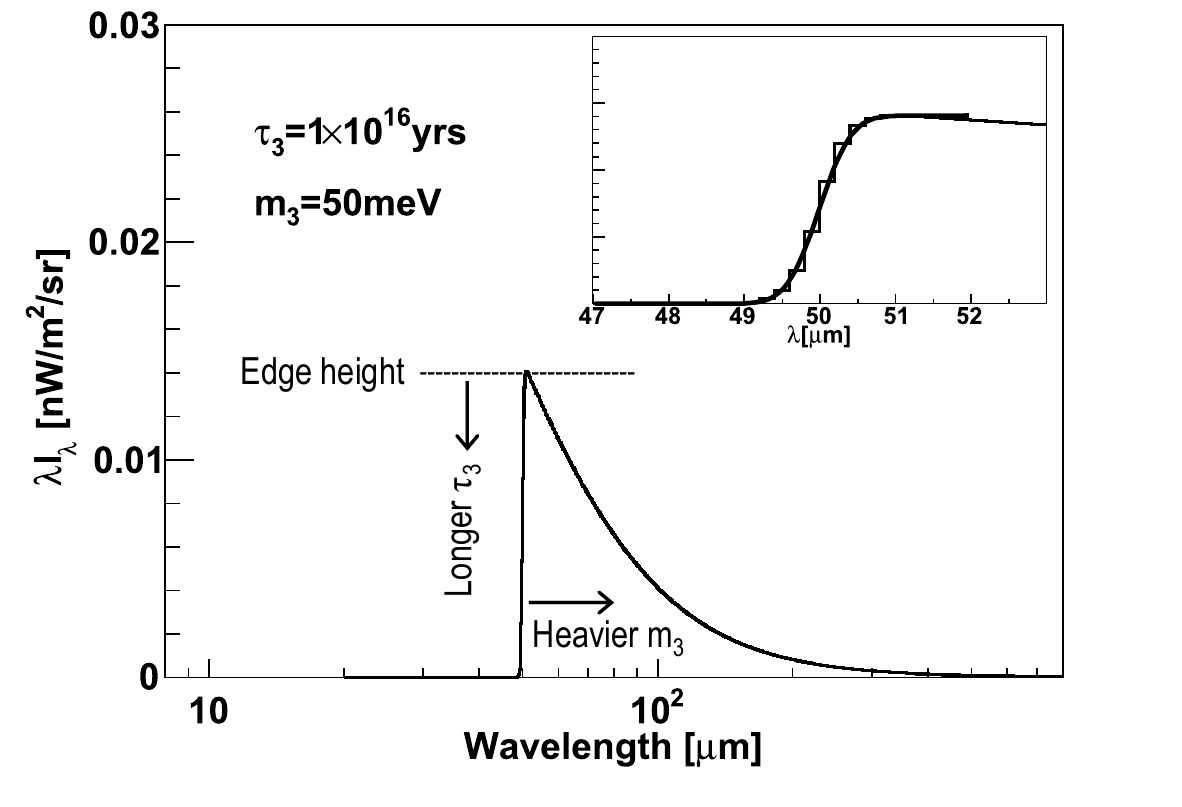}
  \caption{
The expected surface brightness of c$\nu$B decay photons assuming a $\nu_{3}$ lifetime ($\tau_{3}$) of $10^{16}$~years and a mass ($m_{3}$) of 50~meV. The inset on the upper right shows a magnified view of the edge. Arrows indicate the dependencies of the edge position and height: a heavier $m_{3}$ shifts the edge to longer wavelengths, while a longer $\tau_{3}$ lowers the edge height. Note that the shift in the edge position toward longer wavelengths due to a heavier $m_{3}$ also reduces the edge height.
}\label{fig1}
\end{figure}

Figure~\ref{fig1} shows the expected surface brightness distribution of the c$\nu$B decay photons, assuming a $\nu_{3}$ lifetime of $10^{16}$~years and a mass of 50~meV. The characteristic feature of the distribution is a sigmoidal-like shape, with a sharp rising edge approximately 1~$\mu$m in width around the photon wavelength of the decay photon in the rest frame of $\nu_{3}$. The width of this sharp rising edge is related to the current temperature of the c$\nu$B, which is 1.95~K. The surface brightness at the edge, which is inversely proportional to the neutrino lifetime, is predicted to be $14~\rm{pW/m^{2}/sr}$ for \(\tau_{3} = 10^{16}\)~years. The distribution is typical of the photon distribution resulting from the two-body decay of particles with a specific mass uniformly distributed in the universe, and can thus be considered decisive and clear evidence of the c$\nu$B decay. Therefore, detecting the sharp rising edge in the c$\nu$B radiative decay spectrum is important, as it serves as a distinct spectral signature with no known astrophysical sources exhibiting similar characteristics in wavelength space.

\section{Search Strategy for c$\nu$B Radiative Decay and Sensitivity to Lifetime}

The c$\nu$B radiative decay signal is expected to be nearly isotropic, as it originates from neutrinos that are homogeneously distributed throughout the universe. Its characteristic spectral feature is a sharp edge at a specific wavelength, as described in Section~\ref{sec2}. The primary background contributions to this signal are zodiacal emission in the foreground and the cosmic infrared background (CIB). Zodiacal emission, primarily due to thermal radiation from interplanetary dust around 50~$\mu$m, constitutes a significant foreground component, with a surface brightness exceeding the expected c$\nu$B decay signal for $\tau_{3} = 10^{15}$~years by more than three orders of magnitude. Similarly, the CIB, which arises from the cumulative infrared emission of distant galaxies and diffuse extragalactic sources, is expected to be an order of magnitude brighter than the c$\nu$B signal under the same assumption.

Given the overwhelming brightness of these background components, the identification of the characteristic spectral edge of the c$\nu$B decay signal is a particularly effective approach. If these background components consist of continuous spectral components and discrete line emissions, the spectral edge specific to c$\nu$B decay remains distinguishable. To achieve this, our observational strategy involves uniformly scanning a wide region of the sky and accumulating spectral data.

Since zodiacal emission, interstellar dust, and intergalactic dust are all primarily characterized by thermal radiation, their spectral distributions inherently exhibit continuous spectral characteristics. In the case of the CIB, the integrated light from distant galaxies is expected to form a spectrally continuous component when observed over a sufficiently large field of view. This arises due to the redshift distribution of a sufficiently numerous population of distant galaxies, leading to the smearing of emission lines across a broad wavelength range and effectively producing a continuous spectral background.

To maximize the sensitivity to the c$\nu$B radiative decay signal, it is essential to select an observation region where the contributions from zodiacal emission and interstellar dust are minimized. Since zodiacal emission is strongest near the ecliptic plane, selecting a field at high ecliptic latitudes significantly reduces its impact. Likewise, interstellar dust emission is concentrated along the Galactic plane, and its contribution can be mitigated by choosing a region with low dust column density.

A promising choice for such a low-background region is the AKARI Deep Field South (ADF-S)\cite{ref9}, which spans approximately 12 square degrees and is known to have the lowest interstellar dust density in the sky. Additionally, ADF-S is located near the South Ecliptic Pole, ensuring minimal contamination from zodiacal emission. This field has been extensively used for measurements of the CIB and provides a well-characterized galaxy map, facilitating the identification and subtraction of point sources. Although ADF-S is a low-dust region overall, spatial variations in dust density exist within the field. To further optimize the observation, we select a 1-square-degree region within ADF-S that exhibits the lowest interstellar dust contamination. This selection is guided by existing dust maps and infrared observations to minimize background interference and maximize the sensitivity to the spectral edge of the c$\nu$B decay spectrum.

\begin{figure}[htbp]
  \centering
  \includegraphics[width=\columnwidth]{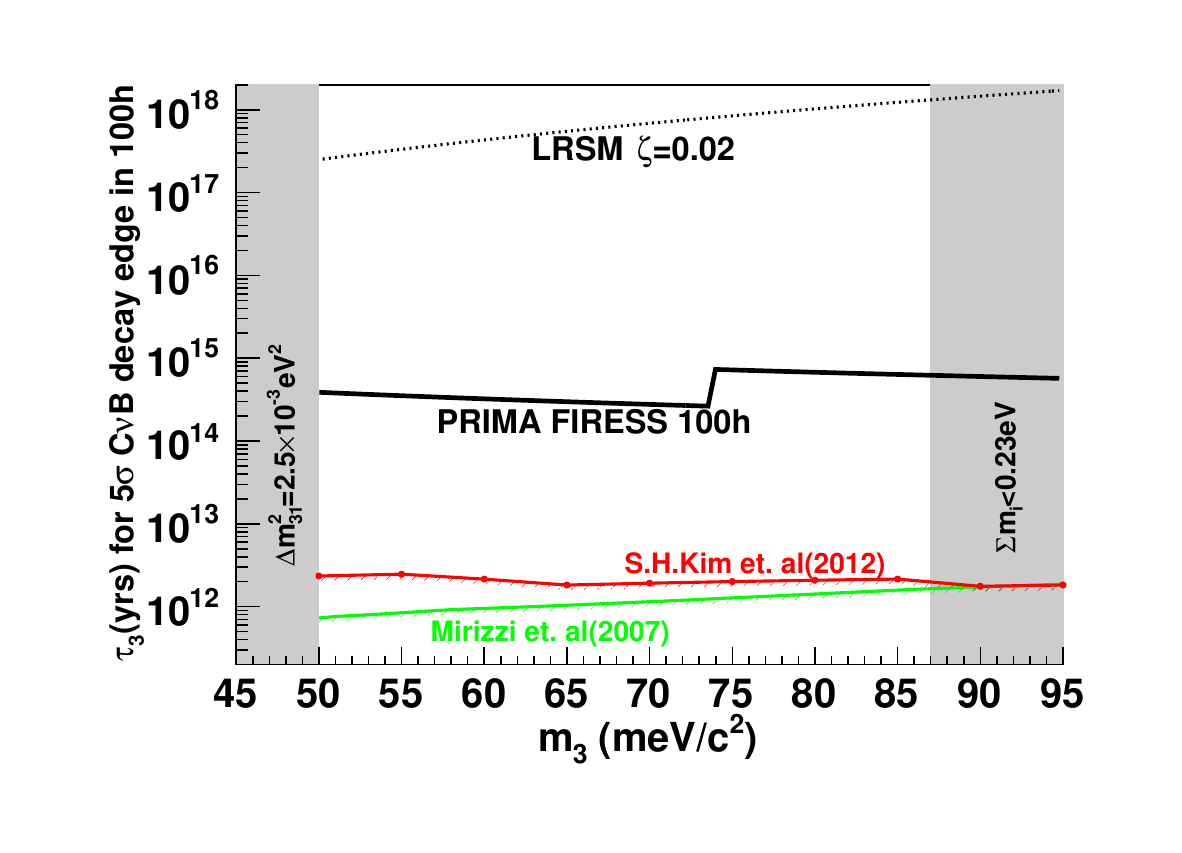}
  \caption{
The mass-dependent estimate of the lifetime of $\nu_{3}$, at which the surface brightness of the edge in the c$\nu$B radiative decay spectrum corresponds to a 5~$\sigma$ sensitivity level obtained from a 100-hour observation. The dashed line represents the predicted $\nu_{3}$ lifetime in the Left-Right Symmetric Model (LRSM), based on the framework discussed in Ref. 5, with a mixing angle $\zeta=0.02$. The current lower limits on the $\nu_{3}$ lifetime obtained from cosmic infrared background measurements, as reported in Refs.~\cite{ref7} and \cite{ref11}, are also plotted.
}\label{fig2}
\end{figure}

To effectively implement this strategy, we utilize PRIMA Far-Infrared Enhanced Survey Spectrometer (FIRESS)\cite{ref10}, a spectroscopic instrument designed for wide-field infrared observations with high sensitivity. The observations will be conducted using the FIRESS low-resolution mode ($R \sim 100$) on the PRIMA space telescope, with a total observation time of 100~hours. FIRESS is equipped with a slit-fed grating spectrometer that covers four spectral bands: Band 1 (24--43~$\mu$m), Band 2 (42--76~$\mu$m), Band 3 (74--134~$\mu$m), and Band 4 (130--235~$\mu$m). Each band has a dedicated detector array with 24 spatial pixels and 84 spectral pixels, and all bands are read out simultaneously. The low-resolution mode ($R \sim 100$) is optimized for high-sensitivity spectral measurements, making it well-suited for detecting the c$\nu$B radiative decay spectrum, particularly in the 50--87~$\mu$m range, where the decay spectrum edge is expected to appear. In FIRESS, Bands 1 and 3, as well as Bands 2 and 4, can simultaneously observe the same target region. However, for compact sources, two-pointing observations are required to fully cover the spectral range. In contrast, when searching for signals like the c$\nu$B decay, which are uniformly distributed across the field of view, this constraint is not necessary, allowing for more efficient use of observation time. This capability is crucial for conducting the proposed 100-hour observation of the c$\nu$B decay spectrum, as it allows for the detection of the spectral edge with sufficient sensitivity and enables constraints on the $\nu_{3}$ lifetime, as presented in Fig.~\ref{fig2}.

In our observation strategy for uniform surface brightness, it can be considered that 24~pixels within the slit field of view are simultaneously measuring a uniformly distributed background over a total of 100~hours. Therefore, using the 5~$\sigma$ flux sensitivity for a 1.2-hour observation of a point source with line emission for FIRESS, the 5~$\sigma$ sensitivity for surface brightness can be expressed by the following equation:

\begin{equation}
\label{eqn4}
5\sigma\left( \lambda I_{\lambda} \right) = \frac{R}{\rm{FoV}_{px}}\sqrt{\frac{1.2~\rm{hr}}{24 \times 100~\rm{hr}}} \times 1.9 \times 10^{- 19}\rm{W/m^{2}}~.
\end{equation}

Here, $\rm{FoV_{px}}$ represents the hexagonal field of view covered by a single pixel. We calculated the $\nu_{3}$ lifetime as a function of the $\nu_{3}$ mass, where the edge formed by the c$\nu$B decay photon spectrum can be detected at a 5~$\sigma$ level --- that is, when the edge height equals the 5~$\sigma$ sensitivity given by Eq.~\eqref{eqn4}.

Figure~\ref{fig2} presents a mass-dependent estimate of the lifetime of $\nu_{3}$, at which the surface brightness of the edge in the c$\nu$B radiative decay spectrum corresponds to a 5~$\sigma$ sensitivity level achievable with a 100-hour observation, together with the current lower limit on the $\nu_{3}$ lifetime obtained from cosmic infrared background measurements, as reported in Refs.~\cite{ref7} and \cite{ref11}. The dashed line represents the predicted $\nu_{3}$ lifetime in the Left-Right Symmetric Model (LRSM), based on the framework discussed in Ref. 5, with the mixing angle $\zeta=0.02$ between the left-handed and right-handed W boson interactions. The estimate of the 5~$\sigma$ sensitivity is obtained by substituting the field of view per pixel of bands 2 and 3, which cover the photon wavelength range of 50--89~$\mu$m (corresponding to 50--87~meV for the neutrino mass $m_{3}$) along with the spectral resolution $R \sim 100$, into Eq.~\eqref{eqn4}. It is indicated that we can explore c$\nu$B decay for lifetimes extending beyond the current lower limit, up to $\rm {O}(10^{15})$~years.

\section{Summary}

In summary, we propose to search for the radiative decay of cosmic neutrino background by observing a region of about 1~square degree for 100~hours. The edge of the spectrum would serve as a clear signal, allowing us to achieve a sensitivity to decay lifetimes up to $\rm {O}(10^{15})$~years, enabled by the uninterrupted spectral coverage from PRIMA around 50~$\mu$m and its deep sensitivity, despite the expected c$\nu$B decay signal at the spectral edge remaining orders of magnitude fainter than the CIB and the zodiacal emission foreground. If neutrino decay is detected with this sensitivity, it would provide compelling evidence for the contribution of non-standard models to neutrino decay, as well as enable the direct detection of cosmic neutrino background. Furthermore, the edge wavelength of the decay spectrum would allow the determination of the absolute mass of neutrinos. Additionally, the long-wavelength tail of the decay spectrum provides a direct probe of the redshift dependence of cosmological parameters, including the matter density and the dark energy equation of state, as described in Eq.~\eqref{eqn2}. In particular, it enables access to the redshift dependence of dark energy density, though extracting precise information requires an accurate understanding of the CIB spectrum. Unlike traditional methods, such as tracing the evolution of large-scale structure, which rely heavily on cosmological models, the decay spectrum tail provides a nearly model-independent probe of the redshift evolution of dark energy. This unique property establishes it as a powerful tool to investigate for the nature of cosmic acceleration.

\subsection{Acknowledgments}

We sincerely thank the organizers of the PRIMA community for their invaluable efforts in fostering collaboration and for providing us with the opportunity to contribute this article.

% references


\begin{thebibliography}{99}

\bibitem{ref1}
  Brent Follin, Lloyd Knox, Marius Millea and Zhen Pan, ``A First Detection of the Acoustic Oscillation Phase Shift Expected from the Cosmic Neutrino Background'', Phys. Rev. Lett. \textbf{115}, 091301 (2015) {[}doi:10.1103/PhysRevLett.115.091301{]}
  
\bibitem{ref2}
  Y. Fukuda et al. (Super-Kamiokande Collaboration), ``Evidence for oscillation of atmospheric neutrinos'', Phys. Rev. Lett. \textbf{81} (8), 1562--1567 (1998) {[}doi:10.1103/PhysRevLett.81.1562{]}

\bibitem{ref3}
  I. Esteban, M.C. Gonzalez-Garcia, M. Maltoni et al., ``NuFit-6.0: updated global analysis of three-flavor neutrino oscillations'', J. High Energ. Phys. \textbf{2024}, 216 (2025) {[}doi.org/10.1007/JHEP12(2024)216{]}

\bibitem{ref4}
  P.B. Pal and L. Wolfenstein, ``Radiative decays of massive neutrinos'', Phys. Rev. D\textbf{25} 766 (1982) {[}doi:10.1103/PhysRevD.25.766{]}

\bibitem{ref5}
  M. A. B. Bég, W. J. Marciano, and M. Ruderman, ``Properties of neutrinos in a class of gauge theories'', Phys. Rev. D\textbf{17}, 1395 (1978) {[}doi:10.1103/PhysRevD.17.1395{]}

\bibitem{ref6}
  A. Aboubrahim, T. Ibrahim, and P. Nath, ``Radiative decays of cosmic background neutrinos in extensions of the MSSM with a vectorlike lepton generation'', Phys. Rev. D \textbf{88}, 013019 (2013) {[}doi:10.1103/PhysRevD.88.013019{]}

\bibitem{ref7}
  S.H. Kim, K. Takemasa, Y. Takeuchi, and S. Matsuura, ``Search for Radiative Decays of Cosmic Background Neutrino using Cosmic Infrared Background Energy Spectrum'', JPSJ \textbf{81} (2012) 024101 {[}doi:10.1143/JPSJ.81.024101{]}

\bibitem{ref8}
  P.A.R. Ade et al. (Planck Collaboration), ``Planck 2013 results. XVI. Cosmological parameters'', A\&A \textbf{571}, A16 (2014) {[}doi:10.1051/0004-6361/201321591{]}

\bibitem{ref9}
  S. Matsuura et al., ``DETECTION OF THE COSMIC FAR-INFRARED BACKGROUND IN AKARI DEEP FIELD SOUTH'', Astrophys. J. \textbf{737} (2011) 2 {[}doi: 10.1088/0004-637X/73{]}

\bibitem{ref10}
  Charles Bradford et al. ``The Far-Infrared Enhanced Survey Spectrometer (FIRESS) for PRIMA: Approach and Estimated Performance,'' (This volume)

\bibitem{ref11}
  A. Mirizzi, D. Montanino and P.D. Serpico, ``Revisiting cosmological bounds on radiative neutrino lifetime'', Phys. Rev. D \textbf{76}, 053007 (2007) {[}doi: 10.1103/PhysRevD.76.053007{]}

\end{thebibliography}
\end{document}